\documentclass{aa}
\usepackage{graphicx}
\newcommand{\etal}{et al.}
\newcommand{\asca}{{\it ASCA}}
\newcommand{\hst}{{\it HST}}
\newcommand{\xmm}{{\it XMM-Newton}}
\def\ergs{erg s$^{-1}$}
\def\flux{erg\,cm$^{-2}$\,s$^{-1}$}
\begin{document}

%%%%%%%%%%%%%%%%%%%%%%%%%%%%%%%%%%%%%%%%%%%%%%%%%%%%%%%%%%%%%%%%%%%%%%%%%
% Title and Abstract Section.
%%%%%%%%%%%%%%%%%%%%%%%%%%%%%%%%%%%%%%%%%%%%%%%%%%%%%%%%%%%%%%%%%%%%%%
\title{The X-ray variability and the near-IR to X--ray spectral energy
distribution of four low luminosity Seyfert 1 galaxies}

\author{I.E. Papadakis\inst{1,2} \and  Z. Ioannou\inst{2,1} \and  W.
Brinkmann\inst{3,2} \and E.M. Xilouris\inst{4}} 
\offprints{I.E. Papadakis;  e-mail: jhep@physics.uoc.gr}

\institute{Physics Department, University of Crete, P.O. Box 2208,
           GR--710 03 Heraklion, Crete, Greece 
\and IESL, Foundation for Research and Technology, 711 10, Heraklion, Crete,
Greece 
\and Max--Planck--Institut f\"ur extraterrestrische Physik,
     Giessenbachstrasse, D-85740 Garching, Germany
\and Institute of Astronomy \& Astrophysics, National Observatory of Athens,
     I. Metaxa \& V. Pavlou, GR--152 36 P. Penteli, Athens, Greece}
     
\date{Received ?/ Accepted ?}

\abstract
{We present the results from a study of the X--ray variability and the near-IR
to X--ray spectral energy distribution of four low-luminosity, Seyfert
1 galaxies.} 
{We compared their variability amplitude and broad band spectrum with
those of more luminous AGN in order to investigate whether accretion in
low-luminosity AGN operates as in their luminous counterparts.}
{We used archival \xmm\ and, in two cases, \asca\ data to estimate their X--ray
variability amplitude and determine their X--ray spectral shape and luminosity. 
We also used archival \hst\ data to measure their optical nuclear
luminosity, and near-IR measurements from the literature, in order to construct
their near-IR to X--ray spectra.}
{The X--ray variability amplitude of the four Seyferts is what one would
expect, given their black hole masses. Their near-IR to X--ray spectrum has the
same shape as the spectrum of quasars which are $10^{2}-10^5$ times more
luminous.}
{The objects in our sample are optically classified as Seyfert 1-1.5. This 
implies that they  host a relatively unobscured AGN-like nucleus. They are also of
low luminosity and accrete at a low rate. They are therefore good
candidates to detect radiation from an inefficient accretion process. However, 
our results suggest that they are similar to AGN which are $10^2 - 10^5$
times more luminous. The combination of  a ``radiative efficient accretion disc
plus an X--ray producing hot corona" may persist at low accretion
rates as well.}

\titlerunning{Low luminosity Seyfert 1 galaxies}

\authorrunning{Papadakis~\etal}

\maketitle
   
%%%%%%%%%%%%%%%%%%%%%%%%%%%%%%%%%%%%%%%%%%%%%%%%%%%%%%%%%%%%%%%%%%%%%%%%
\section{Introduction}
%%%%%%%%%%%%%%%%%%%%%%%%%%%%%%%%%%%%%%%%%%%%%%%%%%%%%%%%%%%%%%%%%%%%%%%%
\smallskip

The current paradigm for active galactic nuclei (AGN) comprises a supermassive
black hole (BH) which is accreting material through a disc. Accretion in  a
geometrically thin, optically thick disc (Shakura \& Sunyaev 1973) is the
favorite mechanism to achieve the high accretion efficiency (of the order of
$\sim 0.1$ or so) which is necessary to explain the large power emitted by AGN.
Theoretical modeling of low accretion rate systems suggests that, when the
accretion rate is below a few percent of the Eddington limit, L$_{\rm Edd}$, AGN
may switch to a different accretion mode characterized by low radiative
efficiency. Most ``radiatively inefficient accretion flow" models (RIAFs; see,
e.g., Yuan (2007) and references therein) suggest that the kinetic  energy
associated with the gas is either advected with the matter into the BH or
redirected into an outflow. 

Many of the nearby AGN are intrinsically faint (exhibiting bolometric
luminosities less than $10^{43-44}$ ergs/s or so) while the central BH in a
large fraction of them is as large as $10^7-10^{8}$ M$_{\odot}$ (see, e.g.,
Panessa \etal\ 2006). Consequently, most of these nuclei accrete at a rate lower
than a few percent of L$_{\rm Edd}$. It is then possible that  accretion in
these objects operates in a radiative  inefficient mode.  For example, according
to Merloni et. al.  (2003), the accretion mode should change below  L$_{\rm
X}$/L$_{\rm Edd} \sim 10^{-3}$ (where L$_{\rm X}$ and L$_{\rm Edd}$ are the
$2-10$ keV and  Eddington luminosity, respectively), while Chiaberge, Capetti \&
Macchetto (2005) suggest that RIAFs may operate in active nuclei with  L$_{\rm
O}$/L$_{\rm Edd} \le 10^{-4}$ (where L$_{\rm O}$ is the optical luminosity in
the $R-$band). However, conclusive evidence for the presence of RIAFs in low
luminosity AGNs (LLAGNs) is still lacking. 

One of the most frequently applied tests to investigate whether a RIAF operates
in LLAGNs is to compare their nuclear spectral energy distribution (SED) in the
near-infrared (IR) to ultraviolet (UV) part of the spectrum with the average SED
of powerful quasars.  At these frequencies  different accretion disc models are
expected to show large differences in spectral shape. RIAFs should lack both the
``big blue bump" and the IR (reprocessed) bump, which instead characterize the
emission from  an optically thick, geometrically thin accretion disc and the
surrounding heated dust. Application of this test to various samples of LLAGNs
has yielded contradictory results. While Maoz (2007) finds that the broad band
SEDs of AGNs with luminosities as low as $\sim 10^{40}$ erg/s are quite similar
to the SED of more luminous  AGN, Ho (1999)  found that the low-luminosity AGN
SEDs have a weak or absent blue bump and are ``radio loud". Quataert, di Matteo
and Narayan (1999) find that the optical/UV spectrum of M81 and NGC~4579
decrease with increasing frequency, in contrast to the ``canonical" quasar
spectrum. Ptak \etal\ (2004) find that NGC~3998 (a ``type 1" LINER galaxy) is
lacking the ``big blue bump", while RIAF models can fit its UV to X--ray SED
reasonably well. Chiaberge \etal\ (2006) also find that the SED of a low
luminosity Seyfert 2 galaxy, namely NGC~4565, is different from that of luminous
quasars.

Another powerful test to identify the presence of RIAFs in low luminosity AGNs
is to compare their  variability properties with the properties of more luminous
AGNs. Ptak \etal\ (1998) found that LLAGNs tend to show little or no significant
short-term variability, despite their small luminosity. This is opposite to what
is observed in powerful AGN, where the variability amplitude increases with
decreasing luminosity. This difference has been interpreted as an indication
that the size of the  X-ray producing region in LLAGNs is significantly  larger
than the size of the X-ray source in ``normal" Seyferts. This is in support of
the hypothesis that advection-dominated accretion operates in LLAGNs, as in this
case one can explain the larger size of the nuclear source. 

Our aim is to compare the X--ray variability amplitude and the optical/X--rays
SED of low luminosity AGN and powerful quasars. We used the Palomar optical
spectroscopic survey of nearby galaxies (Ho, Filippenko \& Sargent 1995) to find
suitable objects for our study. This sample has the advantage of having uniform
and high-quality data that allowed classification of the galactic nuclei to be
determined with well defined and objective criteria (Ho, Filippenko \& Sargent
1997). We only considered  objects which are classified as Seyfert 1 - Seyfert
1.5, because in this case we can be certain of the AGN-like nature of their
nuclear activity, and that the central source is unobscured. There are 11 such
objects in the Palomar sample (Ho \etal\ 1997).  We chose four from these
sources, namely  NGC~4235, NGC~4639, NGC~5033 and NGC~5273, because there exist
estimates of their central black hole mass (hence we can estimate their L$_{\rm
Edd}$) and  their nuclear luminosity is small: their L$_{\rm X}$/L$_{\rm Edd}$
ratio is smaller than $10^{-3}$. Therefore, it is possible that accretion in
these objects operates in a radiatively inefficient mode.

First, we used archival \xmm\ and \asca\ data to measure their X-ray variability
amplitude and we compared it with the variability amplitude of more luminous
Seyferts. Then, by i) fitting  models to the \xmm\ spectra of the four sources,
ii) measuring their nuclear optical luminosity using archival \hst\ data,
and iii) using near-IR  luminosity measurements from the literature, we 
constructed their near-IR to X--ray SEDs and we compared them with the mean,
broad band SED of luminous PG quasars. Our final result that the properties of
the four objects in our sample are similar to those of more luminous AGNs,
should have interesting implications regarding the nature of the accretion mode
in low luminosity active galaxies.  

%%%%%%%%%%%%%%%%%%%%%%%%%%%%%%%%%%%%%%%%%%%%%%%%%%%%%%%%%%%%%%%%%%
\section{The X-ray timing and spectral analysis}
%%%%%%%%%%%%%%%%%%%%%%%%%%%%%%%%%%%%%%%%%%%%%%%%%%%%%%%%%%%%%%%%%%%

\subsection{The \xmm\ observations}

All four Seyferts have been observed by \xmm. Although the exposures are less
than $\sim 15$ ksec long, \xmm\ can provide high quality data for the detection
of short term variations and the accurate determination of their nuclear, X--ray
spectral shape and luminosity.

We have retrieved data from the public \xmm\ archive. Details of the
observations are given in Table~\ref{tab:xmmobs}. The EPIC data were
reprocessed  with the {\small XMMSAS} version 6.5. None of the  observations
were affected by strong background flaring events. With an average  count rate
of a few cts~s$^{-1}$ photon pile-up is negligible for the PN and the MOS
detectors, as verified using  the {\small XMMSAS} task $epatplot$. 

For the PN instrument, source counts were accumulated from a rectangular box of
27$\times$26 {\small RAW} pixels  (1 {\small RAW} pixel $\sim$ 4.1\arcsec)
around  the  position of the source. Background data were extracted from a
similar,  source free, region on the chip.  We selected only single and double
events ({\small PATTERN}$\leq$4, {\small FLAG}$\leq$0; for details of the
instruments see Ehle \etal\ 2007)  in the energy range from 400 eV to 10 keV. 
For the MOS data we  accumulated the source counts using a circular aperture
with  radius 60\arcsec \, centered on the source positions.  The background data
were extracted from a similar region on the same chip. Events with  {\small
PATTERN} $\leq$ 12 and {\small FLAG}=0 were used for the analysis. 

The \xmm\ images for all objects show a bright, compact, unresolved nucleus.
NGC~4639, NGC~5033 and NGC~5273 have also been observed with {\it Chandra}. The
higher resolution {\it Chandra} images reveal the presence of a bright nucleus
in all cases as well. Weak diffuse emission, which is less prominent in the hard
band, appears in the images of NGC~4639 and NGC~5033. There are two off-nuclear
sources (at a distance of less than $\sim 1$ arcmin) in the case of NGC~5033 and
one such source in NGC~5273. These sources are significantly weaker than the
central nucleus in both cases. We therefore  expect that the \xmm\ data should
not be significantly affected by contamination from either unresolved sources or
diffuse emission. 

%---------------------------------- TABLE 1 -----------------------------------------
\setcounter{table}{0}
\begin{table*}[t]
\small
\tabcolsep1ex
\caption{ The \xmm\ observation's details}
\begin{tabular}{lcccccrc}
\noalign{\smallskip} \hline \noalign{\smallskip}
\multicolumn{1}{c}{Object} & 
\multicolumn{1}{c}{Observing date} &
\multicolumn{1}{c}{Instrument} &
\multicolumn{1}{c}{ObsId} &
\multicolumn{1}{c}{Mode} & 
\multicolumn{1}{c}{Filter } &
\multicolumn{1}{c}{Exposure} & 
\multicolumn{1}{c}{Count rate } \\
\multicolumn{1}{c}{ } & 
\multicolumn{1}{c}{ UT } &
\multicolumn{1}{c}{ } &
\multicolumn{1}{c}{ } &
\multicolumn{1}{c}{ } & 
\multicolumn{1}{c}{ } &
\multicolumn{1}{c}{(ksec)} & 
\multicolumn{1}{c}{cts/s$^{(\ddagger)}$ } \\
\noalign{\smallskip} \hline \noalign{\smallskip}
NGC 4235 & Jun. 09, 2004  & PN   & 0204650201 & Large Window  & Thin &  9.6 & 0.96 \\
         &  09:04$-$12:43 & MOS1 &            & Prime Partial & Thin & 10.9 & 0.31 \\
         &                & MOS2 &            & Prime Partial & Thin & 10.9 & 0.31 \\
\noalign{\smallskip}\hline  \noalign{\smallskip} 
NGC 4639 & Dec. 12, 2001 & PN   & 0112551001 & Extd. Full Frame & Medium &  8.8 & 0.28 \\
         & 16:19$-$18:19 & MOS1 &            & Full Window      & Thin   & 14.2 & 0.09 \\
         &               & MOS2 &            & Full Window      & Thin   & 14.2 & 0.09 \\
\noalign{\smallskip}\hline  \noalign{\smallskip}  
NGC 5033 & Dec. 18, 2002 & PN   & 0094360501 & Full Frame  & Medium &  9.1 & 2.40 \\
         & 15:05$-$19:01 & MOS1 &            & Full Window & Medium & 11.5 & 0.68 \\
         &               & MOS2 &            & Full Window & Medium & 11.5 & 0.68 \\
\noalign{\smallskip}\hline  \noalign{\smallskip}
NGC 5273 & Jun. 14, 2002 & PN   & 0112551701 & Extd. Full Frame & Medium &  9.3 & 1.57 \\
         & 13:15$-$16:37 & MOS1 &            & Full Window      & Thin   & 14.2 & 0.57 \\
         &               & MOS2 &            & Full Window      & Thin   & 14.2 & 0.56 \\ 
\noalign{\smallskip}\hline
\label{tab:xmmobs}
\end{tabular}
\medskip

$^{\ddagger}$: Background subtracted values.
\end{table*}
%----------------------------------------------------------------------------

%%%%%%%%%%%%%%%%%%%%%%%%%%%%%%%%%%%%%%%%%%%%%%%%%%%%%%%%%%%%%%%%%%%%%%%%%%
\subsection{Timing analysis}
%%%%%%%%%%%%%%%%%%%%%%%%%%%%%%%%%%%%%%%%%%%%%%%%%%%%%%%%%%%%%%%%%%%%%%%

We created 0.4$-$10 keV band light curves by adding  the MOS1 and MOS2 data
only. Although the PN light curves have a better signal to noise, they are
shorter than the MOS light curves. Since the amplitude of the  X-ray variations
in AGN generally increases with increasing time scale, we decided to use the
longest possible light curves\footnote{In any case, the PN and MOS  light curves
look similar in the time intervals where both instruments were in operation.}.
We binned the combined MOS light curves to intervals of size 500\,s. In this way
the number of photons in each bin is large enough (even in the case of   the
faintest source, NGC~4639) to guarantee the applicability of the traditional
$\chi^2$ test to examine whether a source is variable or not.  We accept that a
light curve shows significant variations if the probability of the ``null
hypothesis" (i.e. the source is not variable) is less than 5\%. 

\begin{figure}
\centering
\includegraphics[height=8.5cm,width=8.5cm]{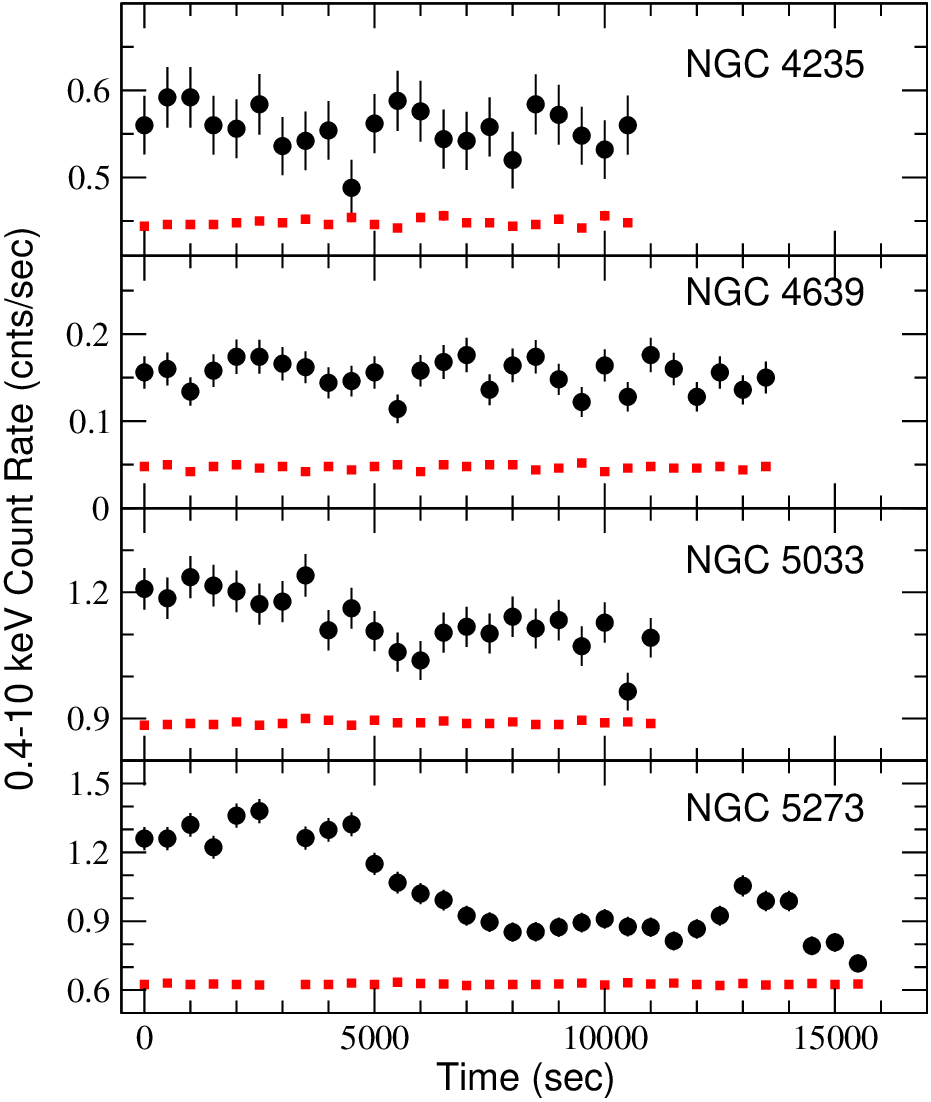}
\caption[]{Background subtracted, combined MOS1+2, 0.4--10 keV light curves of
the four sources, binned in 500\,s intervals. Time is measured from the start of
each observation. Filled squares in all panels indicate the respective
background light curves, rescaled appropriately in each case (errors are smaller
than the symbols size).}
\label{fig:lcs}
\end{figure}

The 0.4$-$10\,keV, MOS1+2 light curves are shown in  Fig.~\ref{fig:lcs}. 
NGC~4235 and NGC~4639 do not show significant variations: $\chi^{2}_{4235} =
15.07/21$ degrees of freedom (dof), and $\chi^{2}_{4639} = 26.1/27$ dof,
respectively. On the other hand we do observe significant variations in the
NGC~5503 and NGC~5273 light curves ($\chi^{2}_{\rm 5033}=44.3/22$ and
$\chi^{2}_{\rm 5273}=552/30$ dof, respectively). 

We use the  normalized excess variance, $\sigma^{2}_{\rm NXS}$ (e.g. Nandra
\etal\ 1997) as a measure of the intrinsic variability amplitude of the light
curves which show significant variations. This estimator is defined as,

\begin{equation}  
\sigma^{2}_{\rm NXS}=\frac{S^{2}-<\sigma_{err}^{2}>}{<x>^2},  
\end{equation}

\noindent where $<x>$, and $S^{2}=(1/\rm {N_{data}})\sum_{i=1}^{\rm
N_{data}}(x_{i}-<x>)^2$ are the mean and variance of the light curve, while 
$<\sigma^2_{\rm{err}}>=(1/{\rm N_{data}})\sum_{i=1}^{{\rm
N_{data}}}\sigma^{2}_{err,i}$ is the average contribution of the Poisson noise
process to the observed scatter around the mean (N$_{\rm data}$ is the number of
points in the light curve). Its square root (the so called ``fractional root
mean square amplitude", $f_{\rm rms}$) indicates the average variability
amplitude of a source as a fraction of the light curve mean. We find that
$f_{\rm rms,5033}=4\pm1$\% and  $f_{\rm rms,5273}=18\pm1$\% (errors account only
for the  measurement errors in the light curve points and have been estimated
according to the prescription of Vaughan \etal, 2003). 

\begin{figure}
\centering
\includegraphics[height=7.5cm,width=8.5cm]{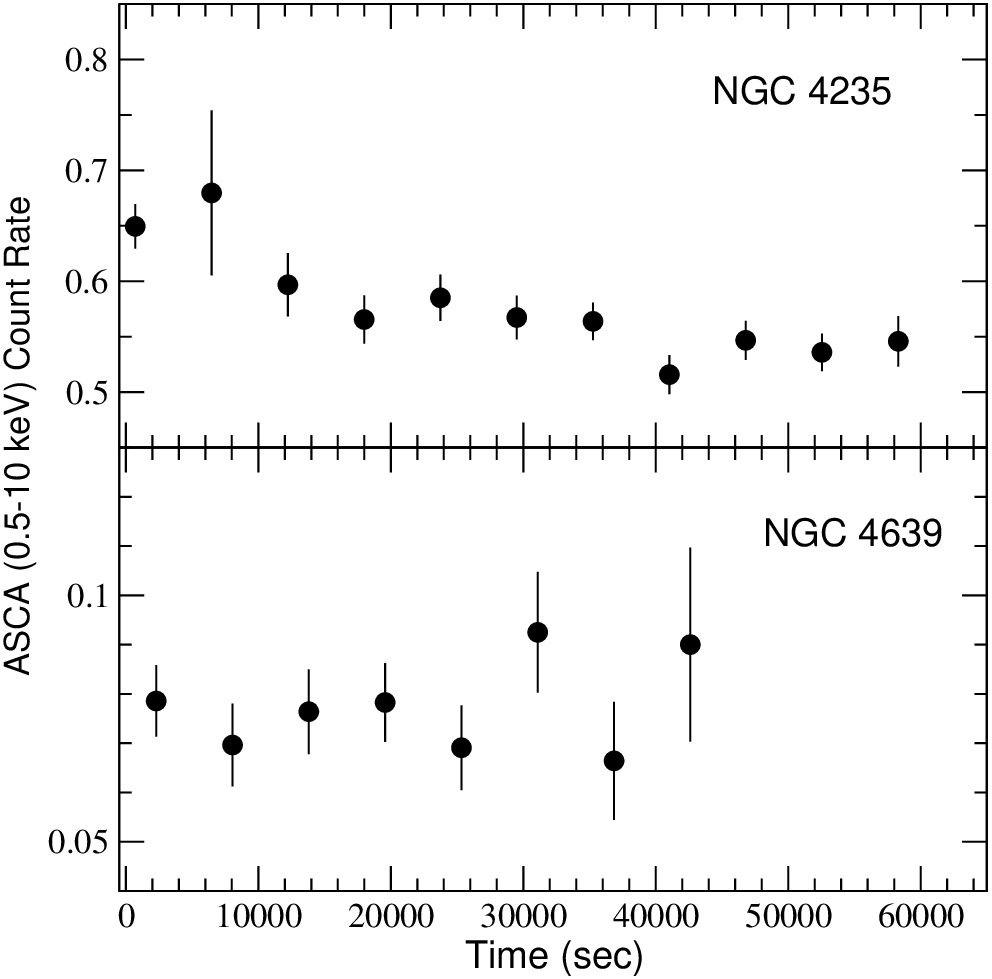}
\caption[]{Background subtracted, combined SIS 0+1, $0.5-10$ keV band light
curves of NGC~4235 and NGC~4639, binned in 5760\,s intervals. Time is measured
from the start of each observation.}
\label{fig:lcs2}
\end{figure}

In order to investigate further the variability of NGC~4235 and NGC~4639, we
considered \asca\ data as well. The main advantage of the  \asca\ light curves is
that they are are  longer than the \xmm\ light curves. In the case of NGC~4235,
we employed its $0.5-10$ keV, 5760 s binned, SIS 0+1 combined  light curves from
the 58 ksec long, December 1998 observation (sequence number 76005000; the light
curves were downloaded from the {\it TARTARUS} database). The background
subtracted light curve is plotted in the upper panel of  Fig.~\ref{fig:lcs2}. 
We detect significant variations ($\chi^{2}=39.8/10$ dof) with $f_{\rm
rms,0.5-10 keV}=6.2\pm 1.8$\%.  

NGC~4639 has also been observed by \asca. However, its background subtracted,
0.5--10 keV, SIS 0+1 combined  {\it TARTARUS} light curve (shown in the bottom
panel of Fig.~~\ref{fig:lcs2}) does not reveal any significant variations during
its $\sim 40$ ksec observation in December 1998 (sequence number 76007000). 

%----------------------------------- FIGURE 1 ---------------------------------
%---------------------------------------------------------------------

%----------------------------------- FIGURE 2 ---------------------------------
%---------------------------------------------------------------------

\begin{figure}
\centering
\includegraphics[height=8.5cm,width=8.5cm]{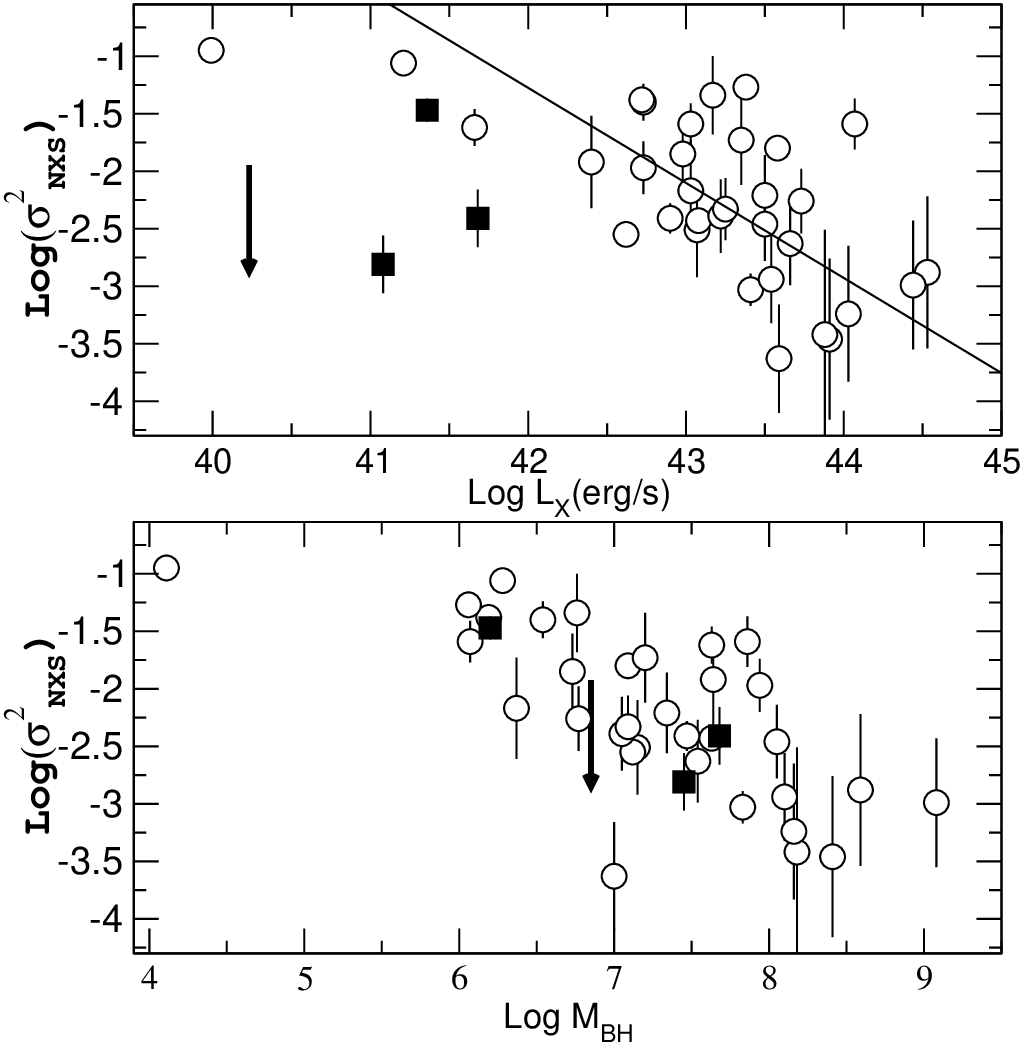}
\caption[]{Top panel:  Log of $\sigma^{2}_{\rm NXS}$ plotted as a function of 
logarithm of X-ray luminosity  for 33 AGN (open circles; data are taken from
O'Neill \etal\ 2005). Filled diamonds indicate the [log($\sigma^2_{\rm NXS}$),
log(L$_{\rm X}$)] data  of  NGC~5273, NGC~5033 and NGC~4235, while the thick
arrow indicates the $3\sigma$ upper limit on  log($\sigma^2_{\rm NXS}$) for
NGC~4639. Bottom panel: The log of $\sigma^{2}_{\rm NXS}$ plotted as a function
of log(BH mass) (in Solar mass units) relation for the 33 AGN of O'Neill \etal\
and the four objects in our sample. }
\label{fig:var}
\end{figure}

The top panel in Fig.~\ref{fig:var} shows the logarithm of $\sigma^{2}_{\rm
NXS}$ plotted as a function of the logarithm of the X-ray (i.e., $2-10$ keV)
luminosity for nearby, bright Seyfert 1 nuclei (open circles; data are taken
from O'Neill \etal\ 2005). The solid line is the best linear fit to the data.
Filled squares in the same figure indicate the  [log($\sigma^2_{\rm NXS}$),
log(L$_{\rm X})$] data of  NGC~5273, NGC~5033 and NGC~4235 (using the X-ray
luminosity estimates listed in Table~3). The thick arrow in  this figure
indicates the upper 3$\sigma$ upper limit of NGC~4639, as estimated using the
MOS1+2 light curves. The plot   suggests that, apart from NGC~5273, the other
three Seyferts are less variable than what we would expect, if we consider the
best line fit to the O'Neill \etal\ data.

However, there has been growing evidence the last few years that the excess
variance in AGN is primarily related to BH mass (Papadakis 2004, O'Neill \etal\
2005). If that is the case, then a ``variability amplitude vs luminosity"
relation is expected only if the objects in the sample are accreting at a
similar rate (Papadakis 2004). In the bottom panel of Fig.~\ref{fig:var}, we
plot the logarithm of $\sigma^{2}_{\rm NXS}$ as a function of the central black
hole mass, M$_{\rm BH}$ (in Solar mass units), for the O'Neill \etal\ objects
and the four Seyferts (BH mass estimates are listed in Table~2). This plot shows
clearly  that their variability amplitudes are in agreement with the amplitudes
we measure in other Seyferts. 

The fact that NGC~5273 shows the largest variability  amplitude among the four
objects in our sample is in agreement with its small black hole mass.  It is not
surprising that the NGC~4235 MOS light curve does not exhibit significant
variations, because it is the shortest of the light curves we study, and this
object has the largest M$_{\rm BH}$ as well. The variations we observe in the
$\sim 5.5$ times longer \asca\ light curve are of the ``correct" amplitude. As
for NGC~4639, it is rather surprising (given its small black hole mass) that
neither the MOS nor the \asca\ light curves reveal any significant variations.
However, this is the faintest among the four sources, so the uncertainty in the
measurement of its $\sigma^{2}_{\rm NXS}$ is quite large. We need a longer \xmm
observation to detect long term variations in this object. 

%----------------------------------- FIGURE 3 ---------------------------------
%---------------------------------------------------------------

Note that, in order to achieve the highest possible signal-to-noise ratio, we
have used  full band light curves (i.e. either $0.4-10$  or $0.5-10$ keV) to
estimate $\sigma^{2}_{\rm NXS}$, while O'Neill \etal\ use $2-10$ keV band light
curves. However, this should not affect our results significantly, as  
$\sigma^{2}_{\rm NXS,2-10 keV}$ and $ \sigma^{2}_{\rm NXS,0.5-2 keV}$ are
comparable in AGN (Nandra \etal, 1997). On the other hand, the  NGC~5053 and
NGC~5273 light curves are half the length of the light curves that  O'Neill
\etal\ use. We therefore conclude that  the X-ray variability mechanism  in
these Seyferts operates in the same way as in more luminous AGN.

%---------------------------------- TABLE 2 ---------------------------------
\begin{center}
\begin{table}
\caption{Distance and BH mass estimates. }
\begin{tabular}{lccc}
\noalign{\smallskip} \hline \noalign{\smallskip}

Object & Distance$^{a}$ & M$_{\rm{BH}}$ & ref$^{b}$ \\
       & (Mpc)          & (M$_\odot$)   &            \\
\noalign{\smallskip} \hline \noalign{\smallskip}

NGC~4235 & $38.0$ &  $4.8\times10^7$ & (1)  \\
\noalign{\smallskip}
  
NGC~4639 & $18.4$ & $7.1\times10^6$ & (2)  \\
\noalign{\smallskip}
   
NGC~5033 & $15.2$ & $2.8\times10^7$ & (3)  \\
\noalign{\smallskip}
    
NGC~5273 & $17.5$ & $1.6\times10^6$ & (4)  \\

\noalign{\smallskip}\hline
\label{tab:objdata}
\end{tabular}

\medskip
$^{a}$ Luminosity distances from NED.\\
$^{b}$ References for the black hole masses: (1) Dong \& de Robertis (2006); 
(2)  Panessa \etal\ (2006); (3) Chiaberge \etal\ (2005); ~~(4) Wu \& Han (2001).
\end{table}
\end{center}
%-----------------------------------------------------------------------

%%%%%%%%%%%%%%%%%%%%%%%%%%%%%%%%%%%%%%%%%%%%%%%%%%%%%%%%%%%%%%%%%%%%%%%%%%
\subsection{X-ray Spectral Analysis}
%%%%%%%%%%%%%%%%%%%%%%%%%%%%%%%%%%%%%%%%%%%%%%%%%%%%%%%%%%%%%%%%%%%%%%%%%%%

For the X-ray spectral analysis, source spectra were grouped with a minimum of 20
counts per energy bin. Spectral fits have been performed with the XSPEC v11.3
package. Spectral responses and the effective area for the PN and MOS spectra
were generated with the {\small SAS} commands {\em rmfgen} and {\em arfgen}. 
Individual fits to the PN and the MOS data resulted in similar model parameter
values. For this reason, we present the results from the joint model fits to the
PN and the MOS data in the 0.4$-$10\,keV energy band. In all model fits we add
Galactic interstellar absorption, using the N$_{\rm H}$  values reported by 
Dickey \& Lockman (1990). The best fitting model parameter values are presented
together with their 90\% errors in Table~\ref{tab:xresults}.  In the same table
we also list our derived 2$-$10 keV fluxes and luminosities.  As K-corrections
can be neglected  for these nearby objects we used the luminosity distances as
given in NED (and listed in Table~\ref{tab:objdata}) for the calculation of
the luminosities. 

%=================================================
\subsubsection{NGC 4639 and NGC~5033}
%=================================================

Cappi \etal\ (2006) have already studied the \xmm\ observations of these two 
sources, and our results are in agreement with theirs. In the case of NGC~4639,
a simple power-law (PL) model with $\Gamma \sim 1.9$ plus cold absorption (with 
N$_{\rm{H}}$ fixed at the Galactic value) resulted in an acceptable fit to the
full band spectra of all three instruments ($\chi^2_{\rm red}$/dof = 1.11/230;
best fitting model values are listed in Table~\ref{tab:xresults}). We do not
find an indication of a cold absorbing column in excess of the  Galactic value,
and the addition of a Gaussian line at $\sim 6.4$ keV does not improve
significantly the goodness of the fit. 

The full band spectrum of NGC~5033 can be fitted well by a power law plus a
narrow Gaussian iron line, assuming galactic absorption only (the line's width
in this case, as well as in the other sources below, is kept fixed at 0.01 keV,
appropriate in the case of a narrow line). The addition of the line improves the
fit quality by $\Delta \chi^2$ = 44.1 for 2 additional dof. The best model
fitting parameters are listed in Table~\ref{tab:xresults}. 

%---------------------------------- TABLE 3 -------------------------------
\begin{table}
%\begin{table*}                                                
%\tabcolsep1ex                                                 
\caption{The X--ray model fit results. } 

\begin{tabular}{lcccc}
\noalign{\smallskip} \hline \noalign{\smallskip} {\smallskip}             
 &   N4235 & N4639 & N5033 & N5273 \\
\noalign{\smallskip} \hline \noalign{\smallskip}
N$_{\rm H,Gal}^\dagger$ & 1.53 & 2.35 & 1 & 0.96  \\
\noalign{\smallskip}
N$_{\rm H,int}^\dagger$ & $16\pm 1$ & -- & -- & $94\pm4$ \\
 \\

{\tt PL} & & & & \\
 $\Gamma$ & 1.61 & 1.86 & 
  1.70 & 1.44 \\
 & ($\pm 0.03$) & ($\pm 0.04$) & ($\pm 0.01$) & ($\pm 0.03$) \\
 A$^\ast$ & $5.9$ & $1.33$ & 
$10.3$ & $10.5^{+0.3}_{-0.5}$ \\
 & ($\pm0.2$) & ($\pm0.04$) & ($\pm0.1$) & \\
\noalign{\smallskip}
\\
{\tt GAUSS} & & & & \\
 E$_l$[keV] & $6.36$       & -- & 
 $6.43$       & $6.40$ \\
 & ($\pm0.02$) &  & ($\pm0.02$) & ($\pm0.02$) \\
 EW[eV]     & 242$^{+124}_{-108}$ & -- & 275$^{+95}_{-85}$ & $190^{+75}_{-65}$ \\
\noalign{\smallskip}
\\
{\tt ABSORI}  & & & & \\
 $\xi^{\rm a}$                        &      --        & -- &        --        & 
 $26\pm$6     \\
\noalign{\smallskip}
\\
{\tt MEKAL} & & & & \\
kT[keV]    & --                   & -- & --                   & $0.18$ \\
 &       & & & $\pm0.03$ \\
\noalign{\smallskip} 
\noalign{\smallskip}	
$\chi^2_{\rm red}$/dof  & 1.08/596 & 1.11/230 & 1.01/901 & 1.06/1042 \\
\noalign{\smallskip}
f$_{\rm 2-10 keV}^{\rm b}$ & 2.8 & 42.7 & 4.3 & 6.3 \\
\noalign{\smallskip}
 L$_{\rm 2-10 keV}^{\rm c}$ & 4.8 & 0.17 & 1.2 & 2.3 \\ 
\noalign{\smallskip}
\hline                                    
\label{tab:xresults}
\end{tabular}
\medskip

\noindent Model fits were performed to the joint PN + MOS1+2 spectra in the
0.4$-$10\,keV range. Errors are at the 90\% level for one interesting
parameter. \newline
$^\dagger$: in units of 10$^{20}$ cm$^{-2}$. \newline
\noindent $^\ast$: PL normalization at 1\ keV in $10^{-4}$ photons cm$^{-2}$ 
 s$^{-1}$ keV$^{-1}$. \newline
\noindent $^{\rm a}$ The ionization parameter which  determines the
absorber ionization state, and is defined as $\xi=L/(nR^2)$. \newline
\noindent $^{\rm b}$ in units of $10^{-12}$\flux . \newline
\noindent $^{\rm c}$ in units of $10^{41}$\ergs . \newline

%\end{table*}
\end{table}
%-----------------------------------------------------------------------

%\vskip 0.2cm \noindent  

%=================================================
\subsubsection{NGC 5273}
%=================================================

The \xmm\ observation of NGC~5273 has also  been studied by  Cappi \etal\ (2006).
For this object our results do not agree with theirs, mainly because they
consider the spectrum over a smaller energy band. A ``PL plus Gaussian line
plus Galactic absorption" model fits well  all the EPIC spectra in the 
2$-$10\,keV band.  When extrapolated  to lower energies, a strong, wide
absorption feature appears around  $\sim$ 1~keV.  A number of complex models
suitable to change the emission characteristics (like absorption edges, partial
covering) resulted in fits of moderate quality ($\chi^2_{\rm red} \leq 1.1$), 
which show systematic deviations from the data in certain energy ranges.

\begin{figure}
\centering
\includegraphics[height=7.5cm,width=8.5cm]{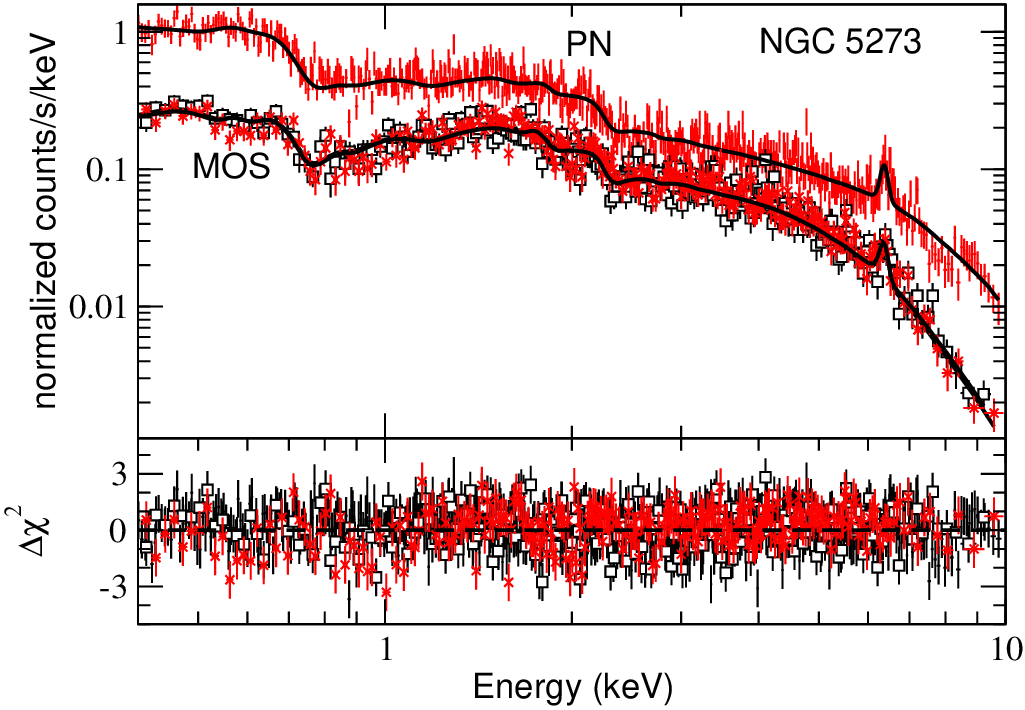}
\caption[]{Simultaneous model fit to the PN, MOS1 (open squares) and  MOS2
(asterisks) fit to the NGC~5273 0.4-10~keV data. The model shown here suggests
that the flux from the central engine and warm circumnuclear material is
absorbed by an ionized absorber, as well as by cold galactic material in the
source. An Fe K line at 6.4 keV can also be seen.}
\label{fig:5273xfits}
\end{figure}

Cappi \etal\ (2006) found that a PL plus a thermal plasma model with $\Gamma\sim
1.4$ and $kT\sim 0.2$ keV can fit well the  spectra. We find that their model
provides a good fit only in the `restricted' energy range of 0.6$-$10\,keV,
while at lower  energies it diverges dramatically from the data. Excellent fits
to all spectra over the 0.4$-$10\,keV energy range could,  finally, be achieved
by a PL plus a Gaussian line plus a thermal plasma emission model (parametrized
with the component {\footnotesize{MEKAL}} in XSPEC), modified by the presence of
an ionized absorber and the cold Galactic material, (i.e. by a {\footnotesize
WABS $\times$ABSORI$\times$ (MEKAL + PL + GAUSS)} model in XSPEC terminology;
the {\footnotesize {ABSORI}} power-law photon index was kept tied to the
{\footnotesize {PL}} slope).This model yielded a $\chi^2$ of  $1.05$ for 1043
dof. The soft thermal flux in the 0.5$-$2\,keV band accounts for about 10\% of
the soft power law flux. Fig.~\ref{fig:5273xfits} shows the  above mentioned
best model-fit to the EPIC data, together with the best  fit residuals. 

%----------------------------------- FIGURE 4 ---------------------------------
%-----------------------------------------------------------------------------

%=================================================
\subsubsection{NGC 4235}
%=================================================

\begin{figure}
\centering
\includegraphics[height=7.5cm,width=8.5cm]{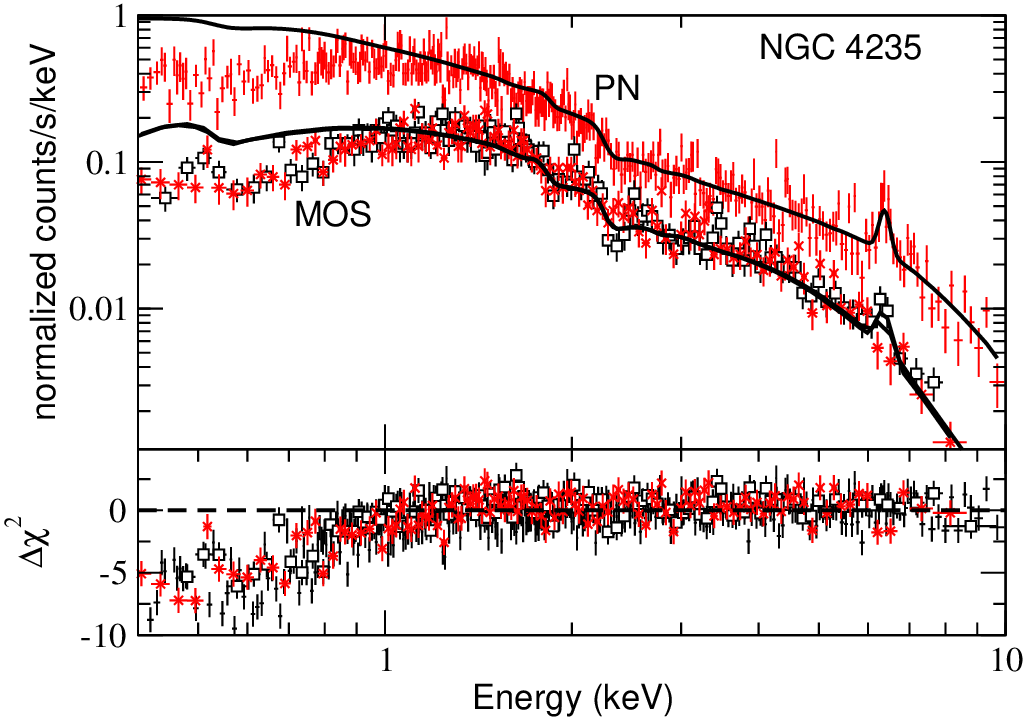}
\caption[]{The 0.4$-$10 keV, EPIC spectra of NGC~4235, plotted together with
the $2-10$ keV, best model fitting ``power-law plus Gaussian line" model (upper
panel). The model is extrapolated to lower energies assuming  Galactic
absorption only. The respective model fitting residuals plot is shown in the
lower panel of the figure (symbols are as in Fig.~\ref{fig:5273xfits}.}
\label{fig:4235xfits}
\end{figure}

The \xmm\ observation of this source has not been studied in the past.  A simple
PL model with ${\Gamma}\sim 1.6$ plus a $\sim 6.4$ Gaussian line fits well the
two MOS and the PN spectra in the $2-10$ keV band. Fig.~\ref{fig:4235xfits}
shows the spectra together with the $2-10$ keV best fitting power law model
extrapolated to low energies, assuming Galactic absorption only (upper panel)
and the respective best model fitting residuals (lower panel). The flux deficit
at energies below $\sim 1$ keV clearly indicates the presence of intrinsic
absorption in this object. 

A PL model plus a Gaussian Line plus a cold absorption component  
({\footnotesize WABS} in XSPEC) with N$_{\rm{H}}$ left as a free parameter 
gives acceptable fits to the full band, EPIC spectra of the source. The best
model fitting results are listed in  Table~\ref{tab:xresults}.  The inclusion of
a Gaussian line improves  the simple PL model fit substantially in the case of
the PN spectrum only ($\Delta \chi^2$ = 22.3 for 2 additional parameters). The
iron line  is not required in the case of the MOS data, because of their lower
statistical quality.

% ---------------------- FIGURE 5 ------------------------------
%
% -----------------------------------------------------------------

%%%%%%%%%%%%%%%%%%%%%%%%%%%%%%%%%%%%%%%%%%%%%%%%%%%%%%%%%
\section{The near-IR to X--ray SEDs}
%%%%%%%%%%%%%%%%%%%%%%%%%%%%%%%%%%%%%%%%%%%%%%%%%%%%%%%%%%%%%%%%%%%%%

In order to construct broad band SEDs we use the results from the spectral
analysis of the \xmm\ data,  nuclear flux measurements from high-resolution
\hst\ images (Section 3.1), as well as near-IR measurements from the literature.

For the near-IR data, we use the nuclear flux measurements of Alonso-Herrero
\etal\ (2003), derived from high-resolution \hst\/NICMOS images at 1.1 $\mu$m
(F110W filter)  and 1.6 $\mu$m (F160W filter). We also use their nuclear flux
measurements in the K ($\lambda_{\rm central}$ = 2.12 $\mu$m) and L-band
($\lambda_{\rm central}$ = 3.51 $\mu$m), derived from  $\sim$ arcsec resolution
images taken by the 3.8 m NASA IRTF telescope on Mauna Kea. The near-IR data
are summarized in Table~\ref{tab:fluxes}. Luminosities at each waveband have been
estimated as described in Section 2.3.

%---------------------------------- TABLE 4 --------------------------
  
\begin{table}
%\begin{table*}
%\small
%\tabcolsep1ex
\caption{ Near-IR and optical-UV fluxes.}
\begin{tabular}{lrrrc}
\noalign{\smallskip}\hline  \noalign{\smallskip}
%\multicolumn{1}{c}{Object} &
%\multicolumn{1}{c}{$\lambda$(\AA)} &
%\multicolumn{1}{c}{Flux$^1$} &
%\multicolumn{1}{c}{log(L$^2$)} &
%\multicolumn{1}{c}{Ref.} \\
Object & $\lambda$(\AA) & Flux$^1$ & log(L$^2$) & Ref. \\
\noalign{\smallskip}\hline  \noalign{\smallskip}
NGC4235  & $3.5~{\mu}$m    & $2.21{\times}10^{-16}$  &  42.13   & 1 \\
         & $2.1~{\mu}$m    & $<4.49{\times}10^{-16}$ & $<42.21$ & 1 \\
         & $1.6~{\mu}$m    & $4.33{\times}10^{-16}$  &  42.08   & 1 \\
         &  6030\AA & $2.33{\times}10^{-16}$  &  41.62   & 2 \\
\noalign{\smallskip}\hline  \noalign{\smallskip}
NGC4639  &  5487\AA & $3.32{\times}10^{-17}$  &  39.87   & 2 \\
         &  4316\AA & $4.50{\times}10^{-17}$  &  39.90   & 2 \\
         &  3004\AA & $2.32{\times}10^{-16}$  &  40.45   & 2 \\
\noalign{\smallskip}\hline  \noalign{\smallskip}
NGC5033  & $3.5~{\mu}$m    & $2.11{\times}10^{-16}$  &  41.31   & 1 \\
         & $2.1~{\mu}$m    & $<5.45{\times}10^{-16}$ & $<41.50$ & 1 \\
         & $1.6~{\mu}$m    & $3.78{\times}10^{-16}$  &  41.22   & 1 \\
         & $1.1~{\mu}$m    & $1.51{\times}10^{-16}$  &  40.66   & 1 \\
         &  6030\AA & $1.04{\times}10^{-16}$  &  40.24   & 2 \\
         &  5487\AA & $1.08{\times}10^{-15}$  &  41.21   & 2 \\
         &  4316\AA & $1.51{\times}10^{-15}$  &  41.26   & 2 \\
\noalign{\smallskip}\hline  \noalign{\smallskip}
NGC5273  & $3.5~{\mu}$m    & $7.35{\times}10^{-17}$  &  40.97   & 1 \\
         & $1.6~{\mu}$m    & $1.96{\times}10^{-16}$  &  41.06   & 1 \\
         & $1.1~{\mu}$m    & $1.69{\times}10^{-16}$  &  40.83   & 1 \\
         &  7898\AA & $1.44{\times}10^{-16}$  &  40.62   & 2 \\
         &  6030\AA & $2.86{\times}10^{-16}$  &  40.80   & 2 \\
         &  5487\AA & $3.60{\times}10^{-16}$  &  40.87   & 2 \\
         &  4316\AA & $5.43{\times}10^{-16}$  &  40.93   & 2 \\
         &  3004\AA & $6.73{\times}10^{-16}$  &  40.87   & 2 \\
\noalign{\smallskip}\hline
\label{tab:fluxes}
\end{tabular}

\medskip
$^{1}$ in \flux \AA$^{-1}$. \\
$^{2}$ in \ergs .\\
References: (1) Alonso-Herrero \etal\ (2003); (2) This work. 
\end{table}
   
Finally, we searched the  MAST archive at STScI for all available WFPC2 images
for the four galaxies, in order to measure their nuclear fluxes in as many as
possible optical wavebands. We describe below the \hst\ data selection and
reduction methods we have used.

%==========================================================================
\subsection{The \hst\ Observations}
%=======================================================================

All four Seyferts have been observed in at least one filter with WFPC2 on board
\hst. A summary of the \hst\ observations we used is given in
Table~\ref{tab:hstobs}. The date of observation, exposure time and \hst\ proposal
number are listed in Columns 2, 4 and 5. Column 3 lists the WFPC2 filters,
together with the respective $\lambda_{\rm central}$ (in parenthesis). For the
galaxies with more than one image in each filter, we chose to study the one with
the largest exposure which did not cause saturation effects in the central
region.

%---------------------------------- TABLE 5 -----------------------------
   
\begin{table}
%\begin{table*}[t]
\small
\caption{ Summary of the \hst\ Observations.}
\begin{tabular}{lcccc}
\noalign{\smallskip} \hline \noalign{\smallskip}
\multicolumn{1}{c}{Object} &
\multicolumn{1}{c}{Observation} &
\multicolumn{1}{c}{Filter} &
\multicolumn{1}{c}{Exp.} &
\multicolumn{1}{c}{Prop.} \\
         & Date          &       & (s) & ID~\# \\
\noalign{\smallskip} \hline \noalign{\smallskip}
NGC4235 & 04/29/1995 & F606W(6030\AA) & 500 & 5479 \\
\noalign{\smallskip}\hline  \noalign{\smallskip}
NGC4639 & 01/15/1995 & F547M(5487\AA) & 230 & 5381 \\
         & 01/15/1995 & F439W(4316\AA)& 700 & 5381 \\
         & 01/15/1995 & F300W(3004\AA) & 1000& 5381 \\
\noalign{\smallskip}\hline  \noalign{\smallskip}
NGC5033 & 05/28/2001  & F606W(6030\AA) & 400 & 8597 \\
         & 12/18/1994  & F547M(5487\AA) & 230 & 5381 \\
         & 12/18/1994  & F439W(4316\AA) & 700 & 5381 \\
\noalign{\smallskip}\hline  \noalign{\smallskip}
NGC5273 & 04/06/1997  & F791W(7898\AA) & 40  & 6419 \\
         & 07/04/2000  & F606W(6030\AA) & 400 & 8597 \\
         & 01/14/1995  & F547M(5487\AA) & 230 & 5381 \\
         & 01/14/1995  & F439W(4316\AA) & 700 & 5381 \\
         & 01/13/1995  & F300W(3004\AA) & 1000& 5381 \\
\noalign{\smallskip}\hline
\label{tab:hstobs}
\end{tabular}
\medskip
\end{table}
%---------------------------------------------------------------------

The WFPC2 images were already  processed by the standard STScI reduction
pipeline (Baggett \etal\ 2002). Since we have chosen images with no saturation
effects, the only additional  processing step required is the removal of cosmic
rays. For that reason, we used the {\it filter/cosmic} task of the MIDAS
astronomical package.  Any residual cosmic ray events as well as bright
foreground stars were removed  by hand. 

In order to measure nuclear fluxes we derived the radial brightness profile of
each galaxy. To this end, we performed ellipse fitting to the isophotes of the
galaxy images.  This choice is motivated by the fact that the isophotes of
galaxies, especially elliptical (E) and lenticular (S0) as well as the bulge of
spiral galaxies, are not far from ellipses. This technique has been widely used
in the past by various authors, mainly as a method of retrieving embedded galaxy
structures that are hidden by the large-scale distribution of light of the main
body of the galaxy (see Xilouris \& Papadakis 2004, and references therein).
Using the {\it fit/ell3} task of MIDAS, which is based on the formulas of Bender
\& M\"{o}llenhoff (1987), we fitted the isophotes of the galaxy with ellipses
and thus derived the average surface brightness profile along the major axis of
the galaxy.  To convert from counts to physical units, we multiplied the pixel
counts by the value of the keyword PHOTFLAM in the image headers and divide by
the value of the EXPTIME keyword.

We also created synthetic PSF's using the {\it Tiny Tim} software (Krist 1995),
which, for WFPC2, produces an accurate representation of the central region of
the PSF, thus appropriate  for our purpose. We aligned the synthetic PSF's with
the position of the nucleus, and constructed the radial profile of the PSF's
with the same method as we do for the galaxies.

\begin{figure}
\centering
\includegraphics[height=7.5cm,width=8.5cm]{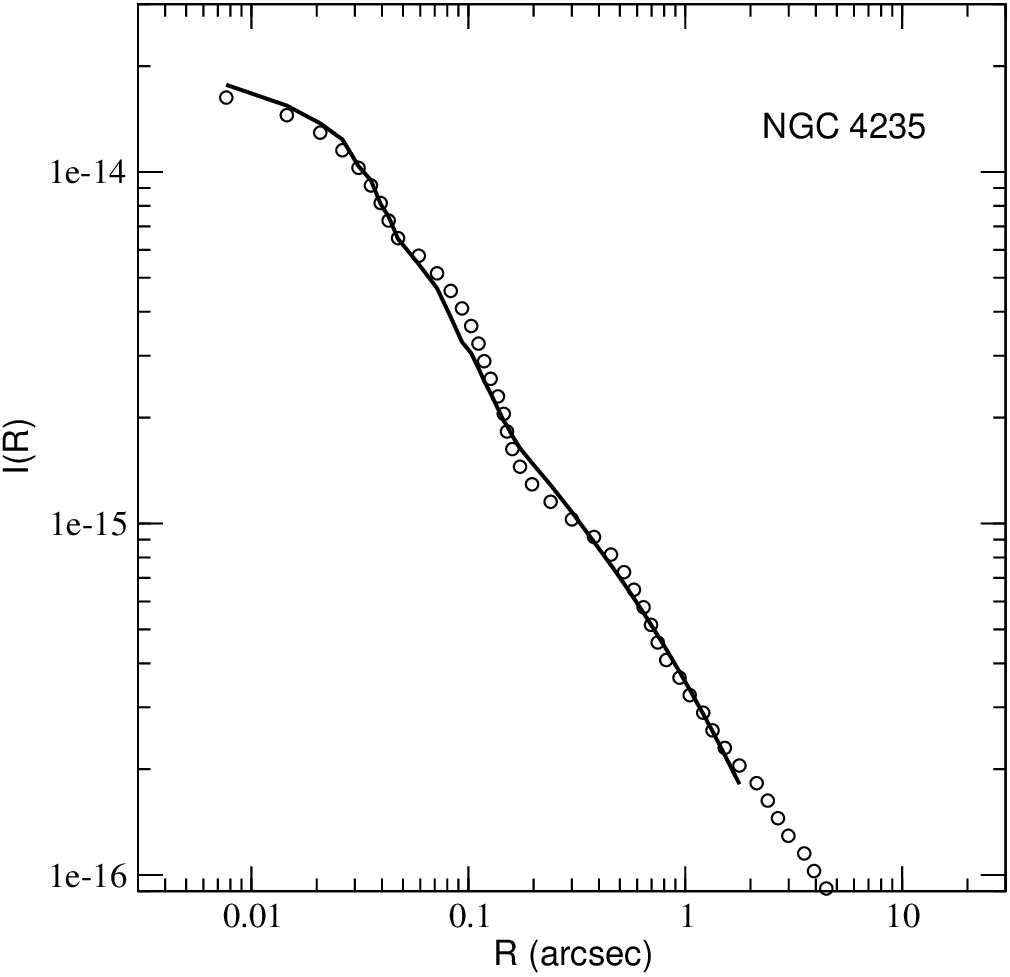}
\caption[]{Radial brightness profile of the central arcsec of NGC~4235 (from the
F6006W image). The solid line shows the best fit to the profile (the fit was
performed to the innermost 2\arcsec\ only)).}
\label{fig:psf}
\end{figure}

Since we are not interested in modeling the bulge of the galaxies on large
scales, we considered the radial brightness profile only in the central
2\arcsec\ and we fitted it with a ``Sersic's function + PSF" model, i.e. a model
of the form  $I(R)=I(0)\exp\{-(R/R_{0})^{1/n}\}+K\times{\rm PSF}(R)$, where
$I(R)$ is the surface brightness, $I(0), R_0,$ and $n$ are constants to be
determined during the fitting process, PSF$(R)$ is the PSF radial profile, and
$K$ is a constant which in effect determines the unresolved nuclear flux.
Fig.~\ref{fig:psf} shows as an example  the brightness profile of the innermost
5\arcsec \, in NGC~4235, together with the best model fitting line. 

%% ----------------------  FIGURE 6 -------------------------------
% -----------------------------------------------------------------

Our results are listed in Table~\ref{tab:fluxes}. The nuclear fluxes are
corrected for Galactic extinction, adopting the values of Schlegel et al.
(1998). In the case of NGC~4235 we also considered intrinsic absorption, given
the clear presence of a  dust lane in its \hst\ image and the X--ray spectral
fitting results which indicate absorption of the nuclear source  by material
within the galaxy. For the reddening correction  in the optical band we
converted the best fitting N$\rm _H$ value to  $A_{V}$ using the relation
$A_V={\rm N_H}/2.2\times 10^{21}$ (Ryter 1996), and we found the extinction at
6030\AA\ using the relations of Cardelli et al. (1989).

Ho \& Peng (2001) have also used the WFPC2/F547M \hst\ images of NGC~4639,
NGC~5033, and NGC~5273 to determine their nuclear flux, using an advanced method
for galaxy image decomposition. They report a nuclear flux of $3.9\times
10^-{17}$,  $9\times 10^-{16}$, and $2.7\times 10^-{16}$ \flux, respectively.
These estimates are in agreement with our results.

%% ----------------------  FIGURE 7 -------------------------------
\begin{figure}
\centering
\includegraphics[height=11.0cm,width=8.5cm]{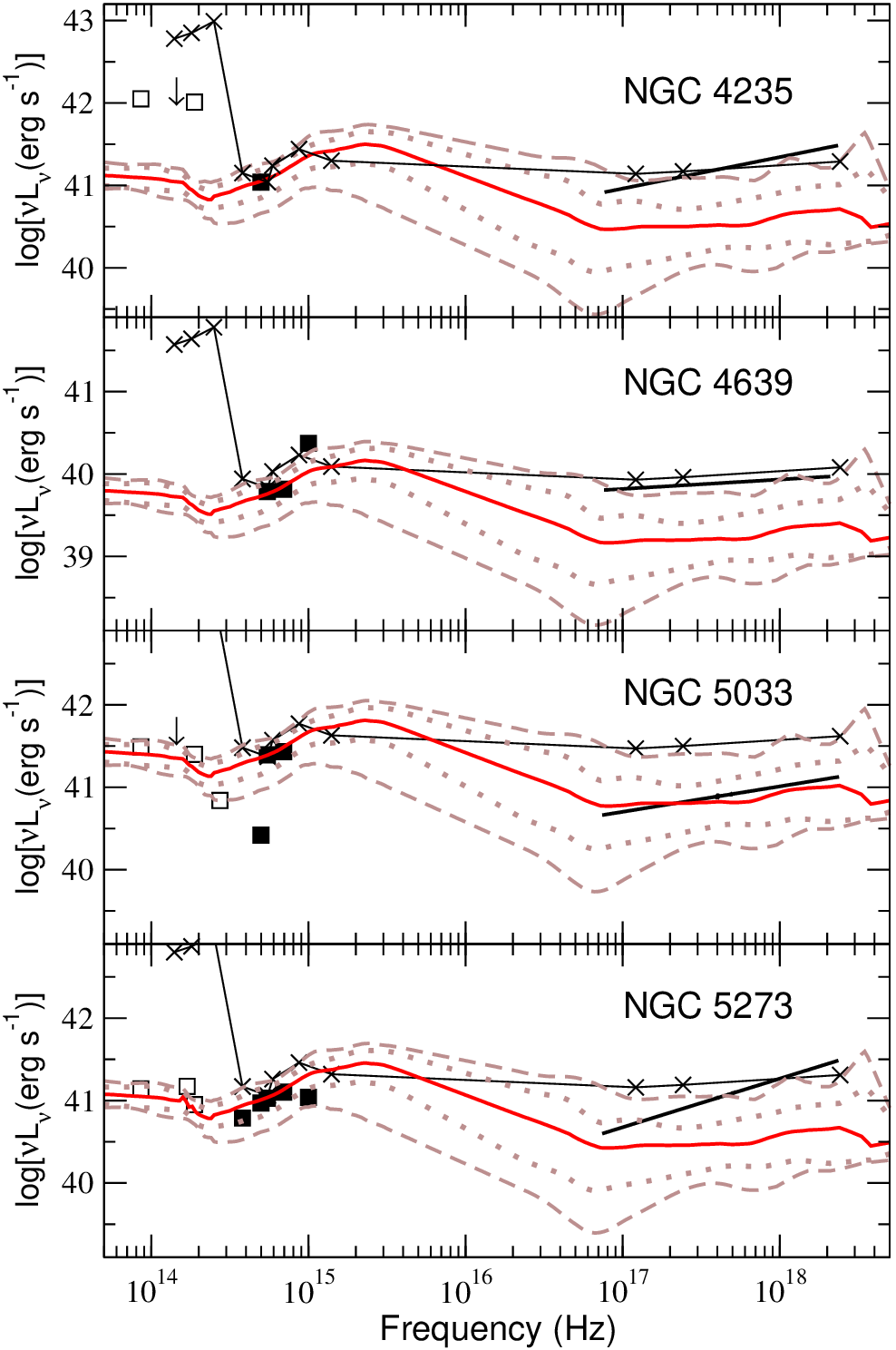}
%\hbox to \hsize{
%\psfig{figure=fig7.eps,height=12.3truecm,width=8.3truecm,angle=0,%
% bbllx=33pt,bblly=41pt,bburx=508pt,bbury=755pt,clip=}
%}
\caption[]{\small The near-IR through the X--ray SED for the four objects in our
sample. Upper limits are 3$\sigma$. The solid, dotted and dashed lines indicate
the mean radio-quiet quasar SED from Elvis \etal, and the the 68 and 90
percentile contours on each side. Filled squares are our measurements from the
\hst\ data, open squares are IR data from the literature. Crosses indicate the
nuclear SED of NGC~3998 (see text fro details).}
\label{fig:seds}
\end{figure}

\subsection{The overall SEDs}

Fig.~\ref{fig:seds} shows the near IR-to-X--ray SEDs of the four Seyferts. The
solid line is the average SED of radio-quiet quasars from Elvis \etal\ (1994);
dotted and dashed lines show the 68 and 90 percentile contours on each side of
the mean SED, respectively. The average quasar SED is normalized to the flux of 
NGC~4235 in the F606W filter, and to the flux of the other three objects in the
F439W filter. Note that all the measurements plotted in Fig.~\ref{fig:seds} are
non-simultaneous. They were taken several years apart, and variations (by
factors of a few, specially in the X--ray band) over these time scales are
common in AGN. This fact could introduce uncertainties which should be kept in
mind when comparing the non-simultaneous measurements with the mean quasar SED.

The near IR/optical SEDs of NGC~5033 and NGC~5273 show a concave spectrum, which
is almost identical, in shape, to the mean quasar SED. This spectral shape  is
often interpreted as the signature of the presence of the ``standard",
geometrically thin/optically thick  disc emission and of dust heated by the
central engine. In NGC~4639 we observe the optical part of the spectrum only,
which is rising towards the UV (as expected if we observe thermal emission from
the accretion disc).

In the case of NGC~4235 and NGC~5033, the X--ray to optical luminosity ratio is
almost identical to what we observe in quasars. In NGC~4639 and NGC~5273, the
optical-UV bump appears to be weaker than the X--rays by a factor of a few.
Since it is in the X--ray band where AGN show their largest amplitude
variations, this difference may not be intrinsic. For example, during its \xmm
observation, NGC~4639 appears to be $\sim 3$ and $\sim 10$  times brighter than
during its \asca (Terashima \etal\ 2002) and {\it CHANDRA} observation (Ho
\etal\ 1999). 

On the other hand, this difference may be intrinsic, indicating that accretion
in these objects operates in a mode which is different than that in powerful
quasars.  To illustrate this point, the crosses in all panels of
Fig.~\ref{fig:seds} show the nuclear SED of NGC~3998. Data were taken from Table
4 of Ptak \etal\ (2004). In the case there were multiple measurements at a
certain wavelength, we chose the ones derived from the smallest aperture size.
The NGC~3998 SED is rescaled to the optical luminosity of the objects in our
sample.  Figure~\ref{fig:seds} shows that the optical to X--ray SEDs of
NGC~3998, NGC~4235 and NGC~4639 (at least) are almost identical.

The NGC~3998 SED is consistent with RIAF models. Ptak \etal\ (2004) found that,
although a ``pure" thin-disc model does not account the mid-IR emission in
NGC~3998, a truncated thin-disc (at around $\sim 300$ Schwarzschild radii),
surrounding an inner RIAF, fitted its mid-IR to X--ray spectrum well (when the
accretion rate is $\sim 10^{-3}$ of the Eddington limit). Consequently,  the
qualitative comparison of the (rather sparse) SEDs of the objects in our sample
with either the average SED of the radio-quiet quasars or SEDs which are
consistent with RIAF models, does not allow us to reach solid conclusions
regarding the accretion mode in these objects. 

\begin{figure}
\centering
\includegraphics[height=8.5cm,width=8.5cm]{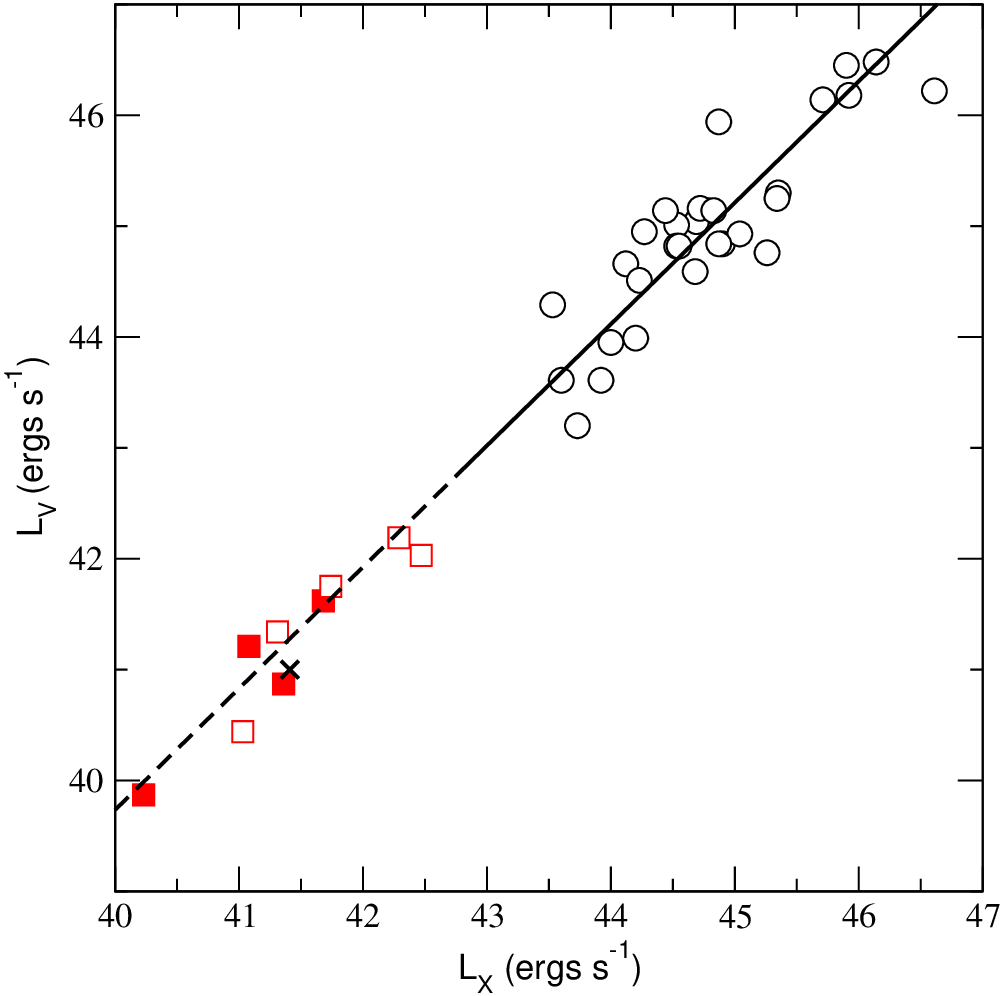}
\caption[]{The L$_{\rm V}$ versus L$_{\rm X}$ plot for PG quasars (open
circles), nearby Seyferts (open squares), the four Seyferts in our sample
(filled squares), and NGC~3998 (cross). The solid line indicates the best linear fit
to the quasar sample, and the dashed line is its extrapolation to lower
luminosities.}
\label{fig:comp}
\end{figure}

We reach more solid results when we compare the optical and X--ray luminosity,
in quasars and LLAGNs, in a more quantitative way. Fig.~\ref{fig:comp} shows the
monochromatic luminosity  L$_{\rm V}$ at 5400\AA, plotted as a function of
L$_{\rm X}$, for PG quasars (data taken from Elvis \etal, 1994).  Filled squares
show the monochromatic luminosity at 5487\AA\ plotted as a function of L$_{\rm
X}$ in the case of the fours Seyferts of our sample (using the data listed in
Tables 3 and 4), while the open squares show data for other nearby Seyferts,
using the nuclear flux  5487\AA\ and X--ray flux measurements of Ho \& Peng
(2001) and Cappi \etal\ (2006).  The cross in the same figure indicates the
5000\AA\ and average 2--10 keV luminosity of NGC~3998, estimated using the flux
measurements of Ptak \etal\ 2004 (as listed in their Table 2) and assuming a
distance of 14.1 Mpc.

We find that a line of the form L$_{\rm V}\propto 1.1(\pm 0.1)$L$_{\rm X}$
(solid line in  Fig.~\ref{fig:comp}) fits the quasar data well. The dashed line
in the same figure shows its  extrapolation to lower luminosities. The agreement
between the low luminosity Seyferts and the powerful PG quasars is excellent. We
do not observe any sign of a ``dichotomy" in this plot, over seven   orders of
magnitude in X--ray luminosity, which might indicate that the low luminosity AGN
accrete in the inefficient radiative regime. In fact, the addition of the low
luminosity objects seems to strengthen the ``optical vs X--ray luminosity"
correlation we detect among the PG quasars alone. We conclude that the optical
to Xray SEDs of the four objects in our sample (and other nearby, low luminosity
Seyferts) is not different from the mean SED of quasars, which are two to five
orders of magnitude more luminous. 

%----------------------------------- FIGURE 8 ------------------------------
%-----------------------------------------------------------------------------

%%%%%%%%%%%%%%%%%%%%%%%%%%%%%%%%%%%%%%%%%%%%%%%%%%%%%%%%%%%%%%%%%%%%%%%%%%%%%%  
\section{Discussion and Conclusions}
%%%%%%%%%%%%%%%%%%%%%%%%%%%%%%%%%%%%%%%%%%%%%%%%%%%%%%%%%%%%%%%%%%%%%%%%%

We have used archival \xmm and \asca data in order to measure the X--ray
variability amplitude of four low luminosity Seyfert 1-1.5 galaxies, and compare
it with the amplitude of brighter AGN. We have used the same  \xmm data to
determine accurately their nuclear  X--ray luminosity and spectral shape, and  
HST data to measure the optical/UV flux of their active nucleus. We then
combined the \xmm spectra, the HST measurements, and near-IR data from
literature to construct  their  broad band, near-IR to X--ray SED and compare
it  with the mean SED of radio-quiet quasars. Our main aim is to  investigate
whether accretion in these objects operates in a mode which is different from
the accretion mode in their more luminous counterparts. 

Such comparisons are frequently used to investigate the nature of the nuclear
source in low luminosity AGNs.  Most of the similar work in the past has focused
mainly  on the comparison between  the properties of galaxies with
low-ionization nuclear emission-line regions (LINERs; Heckman 1980) and luminous
AGN. However, the nature of LINERS and their relation to AGN has been debated
for several years. For this reason, we decided to study objects with a firm
optical classification as a Seyfert 1 - 1.5 galaxy. In this way, we are certain
that they host an accretion driven, AGN-like, UV to X--ray nuclear source. In
addition, the clear presence of broad lines in their optical spectra implies
that we can view their central source directly, and hence determine their
nuclear SED reliably.

Furthermore, using the X--ray luminosities listed in Table~3, and the BH mass
estimates listed in Table~2, we find that the (L$_{\rm X}$/L$_{\rm Edd}$) ratio
of NGC~5273 is $6\times 10^{-4}$. The value of the same ratio in the case of 
NGC~4235, NGC~4639 and NGC~5033 is even smaller: $7.7, 1.8$ and $3.3 \times
10^{-5}$, respectively. This is much smaller than the same ratio of powerful
quasars, and even  20 to 80 times smaller than the ratio of other Seyfert 1
galaxies in the Palomar survey (like for example NGC~4051, NGC~4151 and
NGC~5548). It is also below the limit of $10^{-3}$ where Merloni \etal\ (2003)
suggest that  the accretion mode may change from a thin, optically thick disc to
that of a RIAF. We also converted the measured F547 fluxes of NGC~4639, NGC~5033
and NGC~5273, and the measured F606W flux of NGC~4235, to flux at 7000 \AA,
assuming an optical spectral index of 1, and we estimated the monochromatic
luminosity at this wavelength. Chiaberge \etal\ (2005) suggest that if L$_{\rm
O}$/L$_{\rm Edd}\le 10^{-4}$, then a RIAF may operate in an AGN. In the case of
NGC~4235, NGC~4639 and NGC~5033, this ratio is $6-60 \times 10^{-6}$, while in
the case of NGC~5273,  L$_{\rm O}$/L$_{\rm Edd}\sim1.5\times 10^{-4}$. 

In summary, the four objects of our study  i) should host a relatively
unobscured AGN-like nucleus, ii)  are of low luminosity, and iii) accrete on a
low rate as well. Hence they are good candidates to detect radiation from
RIAF-like processes. 

We detect significant X--ray variations in the \xmm light curves of NGC~5033 and
NGC~5273, and in the \asca light curve of NGC~4235. The  variability amplitude
is 5--15\% of the light curve's mean,  and is  exactly of the ``right" order,
given their BH mass. In agreement with Ptak \etal\ (1998), we find that these
objects do not follow the ``variability amplitude -- luminosity" relation of
other more luminous AGN. However, this result does  not necessarily imply a
larger characteristic size for the X--ray producing region than is the case in
``normal" AGN. On the contrary, if the size is determined by BH mass, as
expected, then our results suggest that, when scaled to BH mass,  the size of
their X--ray sources is similar to the size of the nuclei in more luminous AGN.
Their  ``misplacement" in the ``amplitude -- luminosity" plane is not due to
their variability amplitude being too low but due  to their luminosity being
low,  most probably because of their low accretion rate. Their X--ray spectral
shape is also consistent with the shape of more luminous AGN. Even the best
fitting spectral slope in the case of NGC~5273  is not exceptionally unusual;
NGC~4151, a well studied, luminous Seyfert galaxy, also shows a very flat X--ray
spectrum of $\Gamma\sim 1.5$ (see e.g. de Rosa \etal\ 2007). We therefore
conclude that, from the X--ray  point of view, the central engine in these low
luminosity Seyferts operates in the same way as in the more luminous AGN. 

 The near-IR to X--ray SEDs of the objects in our sample are rather sparse,
and their qualitative comparison with either the average SED of radio-quiet
quasars or the SED of other low-luminosity objects which appear to host a RIAF,
is not conclusive. If we take into account the fact that we used measurements
taken several years apart, and the possibility of large amplitude variations, at
least in the X--ray band, then it is reasonable to conclude that the SEDs of the
AGN in our sample are similar to the average SED of quasars. On the other hand,
the SED of at least NGC~4235 and NGC~4639 are almost identical to the SED of
NGC~3998, which is consistent with a model of a truncated disc and a RIAF at
smaller radii  (see also the case of NGC~4579 and M81 whose SEDs are consistent
with a similar model; Quataert \etal\ 1999). 

The situation becomes clearer when we compare the optical luminosity, L$_{\rm
V}$, with the X--ray luminosity, L$_{\rm X}$, in the 4 Seyferts of our sample
(and a few other LLAGNs) and in quasars (i.e. AGN which are 10$^2$ to $10^5$
times more luminous). We find that the quasar optical luminosity increases
almost proportionally with  L$_{\rm X}$. The extrapolation of the quasar 
``L$_{\rm V}$ vs L$_{\rm X}$" best fit line to lower luminosities agrees very
well with the LLAGN data. It is then reasonable to assume that, the observed
``continuity" in the  (L$_{\rm V}$, L$_{\rm X}$) distribution of quasars and
LLAGNs, also suggests that the optical/X--ray SEDs in AGN are similar, hence,
there is no accretion mode  ``phase transition",  over almost seven orders of
magnitude.

Our results are in agreement with the results of Panessa \etal\ (2006) who study
the X--ray and optical emission line correlation in a sample of AGN with a wide
range of Eddington ratios, and conclude that low-luminosity Seyferts are simply
scaled-down versions of luminous quasars. Similar results were obtained by  Maoz
(2007) who studied the SEDs of 13 low luminosity LINERs, and found that
radiative efficient discs may operate even in these objects, which are
characterized by very low accretion rates.

Although we have tried to use the best available data to determine the SED for
the four Seyferts, the fact that the measurements we use  in the various bands
are separated by years, introduces some uncertainty. Significant improvement can
be achieved if we manage to determine the average nuclear flux of each object 
in various bands. This can be achieved only if we observe individual objects
many times over a time scale of a few years at least, which is not an easy task.
However, the ``variability amplitude test" results  are not affected by these
limitations. Even if we consider only them, the main result of our work is that
accretion operates in the ``normal" way even in objects where the ratio  L$_{\rm
X}$/L$_{\rm Edd}$ is as low as a few$\times 10^{-5}$. 

Perhaps, it is not just the accretion rate that determines the accretion mode in
AGN. Black hole spin and/or mass may also play a role as to how accretion
operates in active galaxies. We plan to investigate this issue by studying the
variability properties of a large sample of low luminosity AGNs, with a wide
range of BH masses. A dichotomy in the ``variability amplitude -- BH mass"
plane, could provide us with important clues as which are the main parameters
which dictate the ``switch" in the AGN accretion mode.   

\vskip 0.4cm
\begin{acknowledgements} This work is based on observations with {\it XMM
Newton}, an ESA science mission with instruments and contributions directly
funded by ESA Member States  and the USA (NASA). It has made use of the
NASA/IPAC Extragalactic Data Base (NED)  which is operated by the Jet Propulsion
Laboratory, California Institute of Technology, under Contract with the National
Aeronautics and Space Administration, and of the Tartarus (Version 3.1)
database, created by Paul O'Neill and Kirpal Nandra at Imperial College London,
and Jane Turner at NASA/GSFC. Tartarus is supported by funding from PPARC, and
NASA grants NAG5-7385 and NAG5-7067. We gratefully acknowledge travel support
through the bilateral Greek-German IKYDA2004 personnel exchange program.    

\end{acknowledgements}

%%%%%%%%%%%%%%%%%%%%%%%%%%%%%%%%%%%%%%%%%%%%%%%%%%%%%%%%%%%%%%%%%%%

\end{document}